\documentclass[prl,twocolumn,amsmath,amssymb,superscriptaddress,showpacs]{revtex4-2}
\usepackage[utf8]{inputenc}
\usepackage[T1]{fontenc}
\usepackage[english]{babel}
\usepackage{graphicx,natbib}
\usepackage{color}

\newcommand\lp{\left(}
\newcommand\rp{\right)}
\newcommand\ls{\left[}
\newcommand\rs{\right]}

\newcommand\la{\left\langle}
\newcommand\ra{\right\rangle}

\begin{document}

\title{Inertial synchronization of networked oscillators in arbitrary dimensions}

\author{Kirill Kovalenko} \thanks{These authors contributed equally to the manuscript.}
\affiliation{Scuola Superiore Meridionale, School for Advanced Studies, Naples, Italy}

\author{Bruce X. Dai}  \thanks{These authors contributed equally to the manuscript.}
\affiliation{Sino-Europe Complex Science Center, North University of China, Shanxi, Taiyuan, 030051, China}

\author{Fanshu Fang}
\affiliation{International Research Center of Complexity Sciences, Hangzhou International Innovation Institute, Beihang University, Hangzhou 311115, China}

\author{Zerong Guo}
\affiliation{School of Physics and Electronic Science, East China Normal University, Shanghai, 200241, China}

\author{Haoran Liu}
\affiliation{School of Mathematics and Statistics, Wuhan University, Wuhan, 430072, China}

\author{Federico Botta}
\affiliation{Department of Computer Science, University of Exeter, North Park Road, Exeter EX44QF, UK}

\author{Charo I. del Genio} \thanks{These authors contributed equally to the manuscript.}
\affiliation{Sino-Europe Complex Science Center, North University of China, Shanxi, Taiyuan, 030051, China}
\affiliation{Institute of Interdisciplinary Intelligent Science, Ningbo University of Technology, Ningbo, China}
\affiliation{Bulgarian Academy of Sciences, Institute of Biophysics and Biomedical Engineering, ul. Acad. Georgi Bonchev 21, Sofia 1113, Bulgaria}

\author{Stefano Boccaletti} \thanks{These authors contributed equally to the manuscript.}
\affiliation{Sino-Europe Complex Science Center, North University of China, Shanxi, Taiyuan, 030051, China}
\affiliation{International Research Center of Complexity Sciences, Hangzhou International Innovation Institute, Beihang University, Hangzhou 311115, China}
\affiliation{Institute of Interdisciplinary Intelligent Science, Ningbo Univ. of Technology, Ningbo, China}
\affiliation{Istituto dei Sistemi Complessi, Consiglio Nazionale
delle Ricerche, via Madonna del Piano 10, I-50019 Sesto Fiorentino, Italy}

\author{Simona Olmi} \thanks{These authors contributed equally to the manuscript.}
\affiliation{Istituto dei Sistemi Complessi, Consiglio Nazionale
delle Ricerche, via Madonna del Piano 10, I-50019 Sesto Fiorentino, Italy}
\affiliation{INFN, Sezione di Firenze, Via Sansone 1, I-50019 Sesto Fiorentino, Italy}

\begin{abstract}
The Kuramoto model provides a paradigmatic framework for studying synchronization
of interacting oscillators, and has been generalized to arbitrary dimensions to describe
swarms, flocks and multi-dimensional opinion dynamics. Yet, existing formulations
neglect inertia, a key mechanism known to enhance information propagation and collective
responsiveness. Here, we introduce and analyze an inertial Kuramoto model in arbitrary
dimensions. We show that inertia fundamentally alters the nature of the synchronization
transition, inducing a crossover from continuous to discontinuous behavior, with
the onset of hysteresis depending explicitly on both inertia and dimensionality parity.
Our analytical theory is supported by extensive numerical simulations. These results
establish inertia as a crucial ingredient of high-dimensional collective dynamics
and reveal a novel universal structure in synchronization phenomena.
\end{abstract}

\maketitle

The emergence of collective dynamics in large populations of interacting
dynamical units is a subject of intense study in physics, sociology, biology
and engineering~\cite{wiesenfeld1998frequency, brunel1999fast, kiss2002emerging, pikovsky2003synchronization, christakis2008collective, luccioli2012collective, vicsek2012collective, witthaut2022collective, Qi25, Sun25, Sem25, Liu25}, with complex networks
providing a suitable representation of the coupling structure of many non-equilibrium
systems.
In this framework, much interest has been devoted to the emergence of nontrivial
collective dynamics in networks of elements whose evolution
is intrinsically simple, such as phase oscillators, whose
state is characterized by a point on a unit circle.

Particular attention has been given to the characterization of synchronized
regimes~\cite{arenas2008synchronization, del15, del16, del22, Jaf24, Par25, Kov26, Guo26},
in which oscillators evolve coherently. Among the paradigmatic frameworks for
investigating synchronization phenomena, the Kuramoto model and its generalizations
play a central role. These models have been successfully applied across a broad
range of systems, from biological contexts, including cardiac pacemaker cells~\cite{osaka2017modified},
flashing fireflies~\cite{ermentrout1991adaptive}, circadian rhythms~\cite{childs2008stability}
and neuronal populations~\cite{breakspear2010generative}, to physical systems,
both natural and engineered, such as power-grid networks~\cite{witthaut2022collective},
superconducting Josephson junction arrays~\cite{marvel2009invariant} and quantum
dipoles~\cite{zhu2015synchronization}. Other classic applications of the Kuramoto
model have been the modelling of the members of a swarm or flock in two dimensions~\cite{sepulchre2004collective, o2017oscillators}
and the characterization of the opinion of individuals in an interacting group~\cite{hong2011kuramoto}.

However, all the aforementioned studies describe collective alignment through a single scalar phase variable, whereas in
many realistic settings interactions occur in higher-dimensional spaces. For instance, velocity alignment in flocks of birds
or schools of fish inherently involves three-dimensional vectors, and opinion dynamics may unfold in spaces of even higher dimensionality,
depending on the range of social attributes considered. Motivated by these considerations, the Kuramoto framework has been generalized
to arbitrary dimensions~\cite{chandra2019continuous}.

Moreover, a fundamental ingredient known to foster information propagation is inertia \cite{cavagna2023natural}, which has been neglected up to now in modelling collective behavior of insect swarms and bird flocks in high-dimensions. Behavioural inertia in the rotations of individual velocities was first introduced to explain the propagation of collective turns in bird flocks  \cite{attanasi2014information, cavagna2015flocking}, but experiments also found clear evidence of underdamped inertial relaxation in natural swarms of midges \cite{cavagna2017dynamic}.


In this article, we bridge this gap by introducing an inertial extension
of the Kuramoto model to arbitrary dimension~$D$. Specifically, we demonstrate
that the nature of the synchronization transition undergoes a crossover
from continuous to discontinuous as inertia increases, with the minimal
inertia required to induce hysteresis that depends explicitly on the dimensionality
of the interaction space. In the vanishing-inertia limit, our framework
naturally recovers the $D$-dimensional Kuramoto model, which emerges as
a special case of the theory. Under reasonable approximations, we provide an analytical characterization
of the transition and corroborate our findings through extensive numerical
simulations.

\begin{figure*}[t]
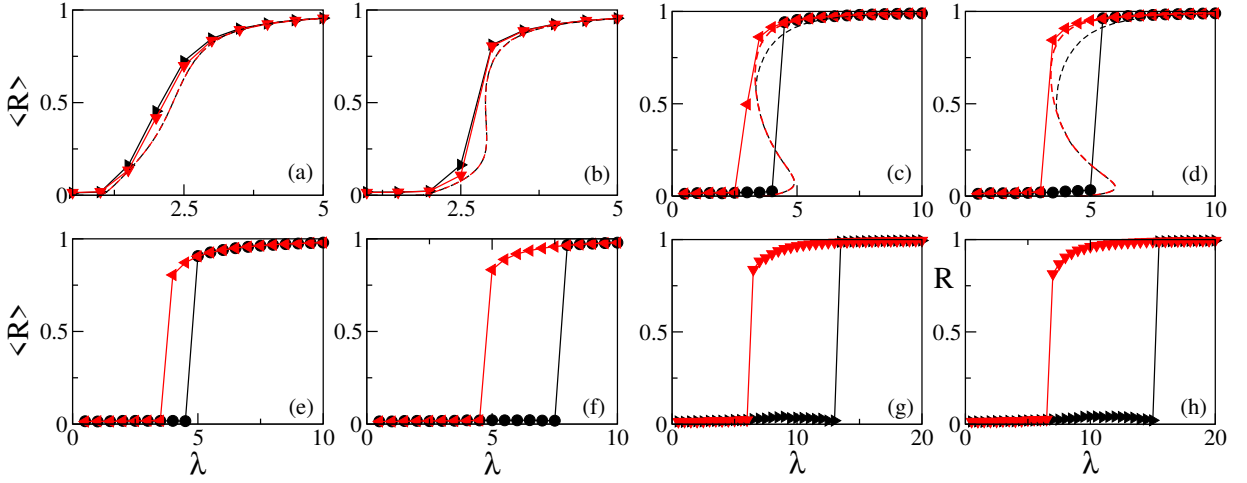

 \includegraphics[width=0.46\textwidth]{Fig1a.eps}
 \includegraphics[width=0.435\textwidth]{Fig1b.eps}
 \caption{\textbf{Change in order of the transition.}
 The time average of the order parameter~$\la R\ra$ vs.\ the
 coupling strength~$\lambda$ shows that the transition
 is continuous for small enough values of~$m$. As $m$
 or the number of dimensions~$D$ increase, a hysteresis
 loop appears. In all panels, the black dots indicate
 the results of forward simulations and the red ones
 those of backwards simulations, the dashed lines with
 corresponding colours are the theoretical predictions,
 and the simulations were carried out for $N=1000$ oscillators,
 for $T_s=100$~time units after a transient time ~$T_t=400$, varying $\Delta\lambda=0.5$. Simulations were performed applying
 a Runge-Kutta $4^{th}$ order integration scheme with integration step 0.001.
 The values of inertia are: (a),(e) $m=0.3$, (b),(f) $m=0.5$,
 (c),(g) $m=0.85$, (d),(h) $m=1$. Panels (a)-(d) refer
 to $D=3$; panels (e)-(h) to $D=5$.}\label{fig:2D3D}
\end{figure*}
We start by considering an ensemble of $N$~rotators, each represented by a $D$-dimensional unit vector $\boldsymbol\sigma_i$ constrained to the surface of a unit $D$-sphere. Here, $\boldsymbol\sigma_i$ represents the orientation of an agent and the evolution equation for the $i$-th rotator (agent) is given by
\begin{multline}\label{DdimKuramoto}
 m\lp\ddot{\boldsymbol\sigma}_i + \left\|\dot{\boldsymbol\sigma}_i\right\|^2\boldsymbol\sigma_i\rp + \dot{\boldsymbol\sigma}_i = \mathbf W_i \boldsymbol\sigma_i \\+ \frac{\lambda}{N} \sum_{j=1}^N \ls\boldsymbol\sigma_j - \lp\boldsymbol\sigma_j \cdot \boldsymbol\sigma_i\rp\boldsymbol{\sigma}_i\rs\:,
\end{multline}
where dots denote temporal derivatives, $m$ is the inertia, $\lambda$ is the coupling strength and $\mathbf W_i$ is a $D\times D$
antisymmetric matrix with elements $\omega_{j,k}^i \sim \mathcal{N}(0,1)$, that represent the natural frequencies of the rotators.
Specifically, the antisymmetric matrix $\mathbf W_i$ describes the intrinsic dynamics of
oscillator $i$ and constitutes the higher-dimensional generalization of
the natural frequency of a phase oscillator. In the absence of coupling,
a constant $W_i$ generates a rigid rotation on the unit sphere through
the evolution operator $e^{W_i t}$. In two dimensions, an antisymmetric
matrix is specified by a single scalar, which coincides with the
conventional natural frequency. Although $\mathbf W_i$ could more generally
depend on time or on $\boldsymbol{\sigma}_i$, while remaining
antisymmetric to preserve $\lVert\boldsymbol{\sigma}_i\rVert=1$, here we
consider constant matrices in order to focus on the effects of
collective interactions and inertia. In the following, while the analytical characterization of the synchronization transition is valid for all natural frequencies distributed according to rotational invariant distribution, numerical simulations are performed by choosing natural frequencies randomly distributed according to a unimodal Gaussian distribution with zero mean and unitary standard deviation. Moreover, the system is initialized with all the velocities tangent to
the sphere, i.e., $\boldsymbol\sigma_i(0) \cdot \dot{\boldsymbol\sigma}_i(0) = 0$.

When the coupling structure among rotators is all-to-all,
Eq.~\eqref{DdimKuramoto} can be conveniently reformulated
in terms of the order parameter $R = \left|\boldsymbol\rho\right| = \left|\frac{1}{N}\sum_{j=1}^N \boldsymbol\sigma_j\right|$
as
\begin{equation}\label{DdimKuramotoMeanField}
m\lp\ddot{\boldsymbol\sigma}_i + \boldsymbol\sigma_i\left|\dot{\boldsymbol\sigma}_i\right|^2\rp + \dot{\boldsymbol\sigma}_i = \mathbf W_i\boldsymbol\sigma_i+\lambda\ls\boldsymbol\rho-\lp\boldsymbol\rho\cdot\boldsymbol\sigma_i\rp\boldsymbol\sigma_i\rs\:,
\end{equation}
which admits a mean-field description for which we can derive the stationary solutions for the order parameter self-consistently.
In this formulation it is easy to see that the coupling term acts as an external torque that tends to align the orientation of each agent with the collective order parameter $\boldsymbol\rho$.

For $D = 2$, this formulation reduces exactly to the standard Kuramoto model with inertia, confirming the model's consistency (see Supplementary Information \cite{supplementary} for the demonstration of the equivalence at $D=2$). For $D=2$, Eq. \ref{DdimKuramotoMeanField} corresponds to a damped driven pendulum equation which admits, for sufﬁciently small forcing frequency $\omega_i$, two ﬁxed points: a stable node and a saddle. At larger frequencies
$|\omega_i| > \Omega_P\approx \frac{4}{\pi}\sqrt{\frac{\lambda R}{m}}$, a homoclinic bifurcation leads
to the emergence of a limit cycle from the saddle. The stable limit cycle and the stable ﬁxed point coexist until a saddle
node bifurcation, taking place at $|\omega_i|= \Omega_D = \lambda R$, leads to
the disappearance of the ﬁxed points, and for $\omega_i>\Omega_D$ only the oscillating solution is present. This scenario is correct for
sufﬁciently large masses, and at small $m$ one has a direct transition from a stable node to a periodic oscillating orbit at
$|\omega_i|= \Omega_D = \lambda R$ \cite{strogatz2024nonlinear}.

As pointed out in Tanaka et al. \cite{tanaka1997first}, the fully incoherent state ($R=0$) and the fully synchronized one ($R\equiv 1$)
correspond to qualitatively distinct dynamical regimes of the oscillators: in the asynchronous state all units drift with finite velocities
$\la\dot{\boldsymbol\sigma}_i\ra$; as the coupling strength increases, oscillators with sufficiently small natural frequencies
$|\omega_i| < \Omega_P\approx \frac{4}{\pi}\sqrt{\frac{\lambda R}{m}}$ begin to lock ($<\dot{\boldsymbol{\sigma}}_i=0>$), while the remaining ones continue
to drift. This progressive locking mechanism ultimately leads to complete synchronization, i.e. $R=1$.
Conversely, when the system is initialized in the fully locked configuration, drifting
motion sets in once the stable fixed-point solution ceases to exist and only a limit-cycle
solution remains. This occurs, upon decreasing the coupling constant, whenever $\left|\omega_i\right|\geqslant \Omega_D = \lambda R$.
In both the cases examined, there is a group of desynchronized oscillators and one of locked
oscillators separated by a frequency, $\Omega_P$ in the first case and $\Omega_D$ in the second one. These groups contribute differently to the total level of synchronization of the system \cite{tanaka1997first}: while drifting oscillators contribute to the total order parameter with a negative contribution, locked oscillators make a positive contribution, with the value of the natural frequency of each unit determining whether and when the unit becomes locked.
In $D$~dimensions, locked oscillators are those aligned with the average orientation
of the population, whereas drifting ones rotate away from it.

\begin{figure}[b]
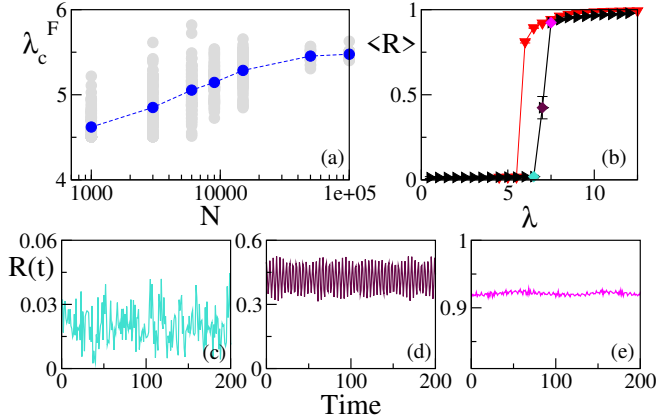

 \includegraphics[width=0.46\textwidth]{Fig2a.eps}
 \includegraphics[width=0.48\textwidth]{Fig2b.eps}
 \caption{\textbf{The high inertia regime.} (a) The
 critical coupling~$\lambda_c^F$ increases with system
 size, eventually reaching approximately~$5.5$. Blue
 dots denote the mean values over different realizations,
 which range from~100 for~$N=1000$ to~20 for~$N=100000$.
 (b) Time average of the order parameter~$\la R\ra$ vs.\ the
 coupling strength~$\lambda$. The black and red triangles
refer to simulations performed by following protocols (I) and (II).
The color code of the filled diamonds corresponds to the color of the time traces
shown in panels (c)-(e), that identify the asynchronous state (turquoise),
cluster state (maroon) and partial synchronization (magenta).
 (c)-(e) Time behavior of the order parameter $R(t)$
 for three different values of~$\lambda$ ($6.5$,
 7 and $7.5$, respectively), exhibiting wide oscillations
 due to the emergence of clusters close to the critical
 coupling in panel~(c). In all panels, the dimensionality
 is $D=3$ and the inertia is $m=5$.}\label{fig:FiniteSize}
\end{figure}
\begin{figure*}[t]
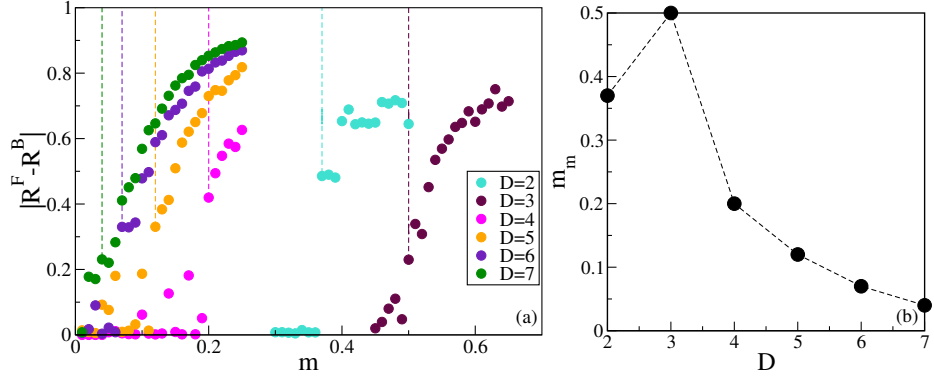

 \includegraphics[width=.394\textwidth]{Fig3a.eps}
 \includegraphics[width=.281\textwidth]{Fig3b.eps}
 \caption{\textbf{The inertia needed to generate a
 first-order transition decreases with the dimensions.}
 (a) Difference between the order parameter values
 in the two transition directions (when positive for
 the first time) during the backwards transition.
 (b) Minimum value of the mass~$m_m$ needed to observe
 a discontinuous transition as a function of~$D$.}\label{fig:Hysteresis-D}
\end{figure*}

Since the distribution of~$\mathbf W_i$ is rotationally
invariant, being multivariate Gaussian, the macroscopic
behavior of the system depends only on the magnitude of~$\boldsymbol\rho$,
and not on its direction. Accordingly, we assume that~$R$
takes a constant value~$\tilde R$, associated with a vector~$\tilde{\boldsymbol\rho}=\tilde R\hat{\boldsymbol\rho}$,
where~$\hat{\boldsymbol\rho}$ is a unit vector. Then, we
derive equations for~$\tilde R$ for the forward and backward
transitions that can be solved self-consistently, determining
the transition curves between the asynchronous state and
the partially synchronized one.

For $D=3$ the transition curves can be derived explicitly.
We briefly outline the derivation here, and refer the reader
to the Supplementary Information \cite{supplementary} for full details. To start,
we make a few assumptions, namely that:
\begin{enumerate}
 \item The distribution of the oscillators is invariant under rotations around~$\tilde{\boldsymbol\rho}$;
 \item The oscillators always switch collectively between a rotating periodic solution and a single stationary
 point, corresponding to the low and high values of the order parameter, respectively;
 \item When the system is in the asynchronous state, the global effect of all interactions is negligible;
 \item When the system is in the asynchronous state, any perturbation dissipates on a time scale that is of
 the same order of magnitude of the order parameter;
 \item For the case of $D=2$, when the system is in the asynchronous state, the average of the rotation speeds of the oscillators is negligible;
 \item In three dimensions, each oscillator is either in a fixed point or is rotating in a fixed plane with almost constant speed.
\end{enumerate}
Then, since~$\mathbf W_i$
is anti-symmetric, for each rotator there exists a basis in
which $\mathbf W_i$ assumes the block-diagonal form
\begin{equation}
\begin{pmatrix}
0 & \omega_i & 0\\
-\omega_i & 0 & 0\\
0 & 0 & 0
\end{pmatrix}\:.
\end{equation}

Projecting Eq.~\ref{DdimKuramotoMeanField} onto the third basis vector yields
\begin{equation}
m\lp\ddot\sigma_{i,3} + \sigma_{i,3} \left|\dot{\boldsymbol\sigma}_i\right|^2\rp + \dot\sigma{i,3} = \lambda\tilde R\ls\hat{\rho}_3 - \lp\hat{\boldsymbol\rho}\cdot\boldsymbol\sigma_i\rp\sigma_{i,3}\rs\:,
\end{equation}
where $(\sigma_{i,1}, \sigma_{i,2}, \sigma_{i,3})$ and $(\hat{\rho}_1, \hat{\rho}_2, \hat{\rho}_3)$
denote the coordinates of $\boldsymbol\sigma_i$ and $\hat{\boldsymbol\rho}$ in this basis.
For a fixed~$\hat{\boldsymbol\rho}$, Eq.~\ref{DdimKuramotoMeanField} admits a solution with constant~$\sigma_{i,3}$ satisfying
\begin{equation}\label{eq:fixedpoint}
\hat{\rho}_3 = \ls\hat{\boldsymbol\rho}\cdot\boldsymbol\sigma_i + \frac{m}{\lambda\tilde R}\omega_i^2\lp 1-\sigma_{i,3}^2\rp\rs\sigma_{i,3}\:,
\end{equation}
which characterizes the rotating solution in the transformed reference frame~\cite{tanaka1997first}.

\begin{figure*}[t]
 \includegraphics[width=0.32\textwidth]{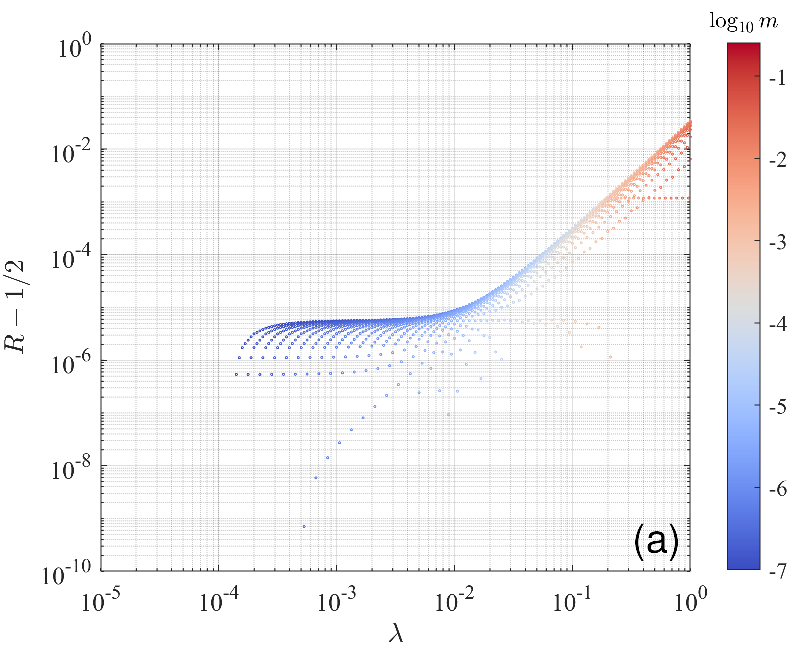}
 \includegraphics[width=0.35\textwidth]{Fig4b.eps}
 \caption{\textbf{The low inertia regime: recovering
 the classic $D$-dimensional Kuramoto model.} (a)
 For vanishing $\lambda$, the order parameter approaches
 $R=\frac{1}{2}$ for small-enough masses. (b) The
 scaling exponent of the smallest coupling~$\lambda_c^B$
 for which partial synchronization is observed with
 the mass crosses over from approximately~$0.5$ to
 approximately~1 as the mass decreases.}\label{fig:BackTransitionOddD}
\end{figure*}
To determine the transition points, Eq.~\eqref{eq:fixedpoint} is solved to identify the value
of~$\sigma_{i,3}$ corresponding to the quasi-stationary third component of the periodic rotating
solution. One then assesses whether the rotating motion in the first two coordinates constitutes
a unique stable periodic solution of the rescaled system. For the backward transition, if this
condition holds, the average projection of~$\boldsymbol\sigma_i$ onto~$\tilde{\boldsymbol\rho}$
reads
\begin{equation}
 \hat{\boldsymbol\rho}\cdot\boldsymbol\sigma_i = -\frac{1}{2} \lambda m \tilde R\frac{r^2}{\lp m\left|\omega_i\right|\rp^3} + \hat\rho_3\sigma_{i,3}\:,
\end{equation}
where $r$ denotes the norm of the subvector formed by the first two components of~$\hat\rho$.
If instead a stable fixed point exists, the contribution becomes
\begin{equation}\label{fixed}
\hat{\boldsymbol\rho}\cdot\boldsymbol\sigma_i = r\sqrt{1-\sigma_{i,3}^2}\sqrt{1-\lp\frac{\omega_i^2}{\lambda\tilde R}\rp^2}\:,
\end{equation}
which corresponds to the stationary fixed-point solution.
The expected value of the order parameter along the backward
threshold is then obtained by integrating the projection
over the distribution of natural frequencies, accounting
for all oscillators, and solving the resulting equation self-consistently.

Similarly, for the forward transition,
if the stability criterion holds, the
average projection of~$\boldsymbol\sigma$
onto~$\tilde{\boldsymbol\rho}$ is
\begin{equation}
 \hat{\boldsymbol\rho}\cdot\boldsymbol\sigma_i = -\frac{1}{2}\frac{\lambda m\tilde Rr^2}{1+\lp m\omega_i\rp^2} + \hat\rho_3\sigma_{i,3}\:;
\end{equation}
otherwise, Eq.~\eqref{fixed} holds. Again,
the order parameter is obtained by integrating
the projection and solving the resulting equation
in a self-consistent way.

To validate the analytical predictions, we performed extensive numerical
simulations by varying the coupling strength~$\lambda$ adiabatically. For
the forward transition, simulations were initialized from random configurations,
and~$\lambda$ was increased from~0 up to a maximum value~$\lambda_M$ in
increments of~$\Delta\lambda$. Subsequently, the protocol was reversed,
decreasing~$\lambda$ stepwise by~$\Delta\lambda$ until returning to~0. Each
simulation, except for the first one, was initialized from the final state of
the preceding run. For every value of~$\lambda$, the system was evolved
over a transient time~$T_t$, followed by an additional interval~$T_s$ during
which the time-averaged value of~$R$ was computed.

The results, reported in Fig.~\ref{fig:2D3D}, demonstrate
that the synchronization transition is continuous for small
inertia and becomes hysteretic beyond a critical inertia
threshold. In particular, in Fig. ~\ref{fig:2D3D} are reported the cases D = 3 (panels (a)-(d))
and D = 5 (panels (e)-(h)). However, this phenomenology is robust across dimensions
from~2 to~7, and consistent with previous findings~\cite{olmi2015chimera, laing2019dynamics}.
The theoretical curves (dashed lines) reported in panels (a)-(d) exhibit excellent agreement with numerical
simulations in the discontinuous jumps and shape of the coherent branch, confirming the validity of the self-consistent
analysis and accurately capturing the dependence of the transition on ~$m$. Also, for fixed~$m$, the width of the
hysteresis loop increases with~$D$. It is important to notice that the theoretical curves for the forward and backward transitions are predicted through an extension of the Kuramoto self-consistent theory for a second order system with dimension $D=3$.
The theory is able to predict the discontinuity of the transition with respect to the coupling strength regardless
of the fact that it approximates a large but still finite-size system \eqref{DdimKuramoto} by the continuum (infinite size)
limit. However, the self-consistent equations for the order parameter are not necessary a good approximations in the vanishing
coupling strength limit, as already shown in \cite{tanaka1997first} for $D=2$, and the end point of the lower backward branch
is not captured properly since the self-consistent theory does not provide the instability properties of the incoherent state.
For larger $m$ ($m>1$), although the decreasing backward branch shows good agreement to the theoretical prediction, the
forward branch generally shows a growing deviation from the theoretical prediction due to the break of
the independence between the whirling oscillators, and the emergence of a secondary synchronization of groups
of the whirling oscillators with lower natural frequencies (drifting clusters).

Additionally, the synchronization transition exhibits finite-size effects,
as illustrated in Fig.~\ref{fig:FiniteSize}(a). In particular, the forward
critical coupling~$\lambda_c^F$ increases with the system size and saturates
to an asymptotic value $\lambda_c^F \approx 5.5$, in agreement with established
results for $D=2$~\cite{olmi_hysteretic_2014}. As mentioned above, and similarly to
what is known for~$D=2$~\cite{tanaka1997first, olmi_hysteretic_2014, olmi2016dynamics}, sufficiently
large inertia leads to the formation of drifting clusters with finite average
velocities within the partially synchronized regime. Furthermore,
the transition between the asynchronous state, shown in Fig.~\ref{fig:FiniteSize}(c),
and the synchronized one, shown in Fig.~\ref{fig:FiniteSize}(e),
proceeds through an intermediate cluster state characterized by pronounced oscillations of the order parameter,
shown in Fig.~\ref{fig:FiniteSize}(d).
These oscillations originate from large secondary clusters undergoing coherent
rotation with finite velocities. Remarkably, the emergence of such macroscopic
oscillations is anomalous: while typically associated with critical (continuous)
transitions, here they arise in a discontinuous, first-order synchronization
transition.

To investigate the dependence of hysteresis on inertia and dimensionality,
we performed systematic simulation cycles in which the coupling strength~$\lambda$
was first increased until the order parameter reached~$0.8$, and subsequently
decreased back to~0. For each value of~$\lambda$, we evaluated the difference
between the forward and backward branches of the order parameter, $R^F$ and~$R^B$,
identifying hysteresis whenever $\left|R^F-R^B\right|>0$ along the transition.
Repeating this protocol for different masses~$m$, we determined,
for each dimension~$D$, the minimum mass~$m_m$ required to induce a first-order transition.
Specifically, we define $m_m$ as the smallest value of~$m$ such that $|R^F-R^B|>0.2$.
As shown in Fig.~\ref{fig:Hysteresis-D}, the minimum inertia necessary for
hysteresis decreases systematically with increasing dimension, with the notable
exception of $D=3$. The non-monotonic behavior observed in panel (b) can be explained
on the basis of the emergence of drifting clusters during the forward transition to synchronization, which is present only
for D=2, 3. The presence (absence) of intermediate steps in the forward branch can result in an increasing (decreasing) dependence
on the inertia for increasing dimensionality.

Finally, we analyzed the backward transition in the limit of vanishing inertia.
While in even dimensions, the system reaches the asynchronous state with vanishing $R$ in a continuous manner independently on the inertia value, for odd dimensions the transition is more complex, and for vanishing inertia the system must recover the behavior of the standard high-dimensional Kuramoto model, whose transition is discontinuous at zero coupling where $R$ reaches $1/2$ \cite{chandra2019continuous, dai_discontinuous_2020}.
More specifically, the system approaches the asynchronous state through a continuous transition in even dimensions.
In contrast, odd dimensions exhibit a qualitatively different behavior: the transition is discontinuous in the limit of vanishing inertia value, as shown in Fig. \ref{fig:BackTransitionOddD}. As illustrated in panel (a), the order parameter converges
to $R=\frac{1}{2}$ as both~$\lambda$ and~$m$ approach zero. Moreover, in the
vanishing-inertia regime, the minimum coupling~$\lambda_c^B$ required to sustain
partial synchronization scales approximately linearly with~$m$ (see panel (b)).
Note that for non-vanishing, finite masses, the system is not able to reach $R=1/2$ at $\lambda=0$ and the order parameter remains finite for $\lambda_c^B>0$, scaling approximately with $\sqrt{m}$.

In conclusion, we generalized the Kuramoto model with inertia to arbitrary dimensions, describing a system of interacting, orientable
rotators, whose state is completely described by D-dimensional unit vectors.
We showed that inertia fundamentally reshapes the macroscopic dynamics of the system:
the synchronization transition is continuous at low inertia and becomes discontinuous
beyond a dimension-dependent threshold. In general, the inertia required to induce a first-order
transition decreases with increasing dimensionality~$D$. In the limit of vanishing inertia,
our framework recovers the $D$-dimensional Kuramoto model, including its characteristic
dependence on dimensional parity. Furthermore, we derived analytical predictions for both
forward and backward synchronization transitions for the case $D=3$ and validated them through large-scale
numerical simulations. Altogether, our results establish inertia as a key control parameter
of high-dimensional collective dynamics and provide a unified theoretical framework for
synchronization phenomena beyond the phase-only description.

\begin{acknowledgments}
This work is partly supported by the Italian PRIN research project n.2022FHHHPC titled ``The structure, dynamics and control of network systems with higher-order interactions''.
Moreover, S.B. acknowledges support from the project n.PGR01177 of the Italian Ministry of Foreign Affairs and International Cooperation.
\end{acknowledgments}

%

\end{document}